\documentclass[pdflatex,sn-mathphys-ay]{sn-jnl}

\usepackage{hyperref}
\usepackage{braket}
\usepackage[inline, shortlabels]{enumitem}

\let\orcidlogo\relax 
\usepackage{orcidlink}
\usepackage{graphicx}%
\usepackage{multirow}%
\usepackage{amsmath,amssymb,amsfonts}%
\usepackage{amsthm}%
\usepackage{mathalpha}
\usepackage[title]{appendix}%
\usepackage{xcolor}%
\usepackage{textcomp}%
\usepackage{manyfoot}%
\usepackage{booktabs}%
\usepackage{algorithm}%
\usepackage{algorithmicx}%
\usepackage{algpseudocode}%
\usepackage{listings}%
\theoremstyle{thmstyleone}%
\theoremstyle{thmstyletwo}%

\theoremstyle{thmstylethree}%

\begin{document}

\title[Article Title]{Divide et Impera quantum neural networks for modular hybrid computing architectures}


\author*[1]{\orcidlink{0000-0003-3840-3609} \fnm{Emanuele} \sur{Casciaro}} \email{emanuele.casciaro@unifi.it}

\author[2]{\orcidlink{0009-0008-6980-7095} \fnm{Fabio} \sur{Mascherpa}} \email{fabio.mascherpa@eni.com}

\author[2]{\orcidlink{0000-0002-7473-5945}~\fnm{Alfonso} \sur{Amendola}} \email{alfonso.amendola@eni.com}

\author[1]{\orcidlink{0000-0002-8366-4296} \fnm{Filippo} \sur{Caruso}} \email{filippo.caruso@unifi.it}

\affil[1]{\orgdiv{Department of Physics and Astrophysics}, \orgname{University of Florence}, \orgaddress{\street{Via Sansone, 1}, \city{Sesto Fiorentino}, \postcode{50019}, \country{Italy}}}

\affil[2]{\orgdiv{DICOX/C – High Performance Computing Center of Excellence, DIT – Digital \& Information Technology}, \orgname{Eni S.p.A.}, \orgaddress{\street{ Via Emilia 1}, \city{San Donato Milanese}, \postcode{20097}, \country{Italy}}}

\abstract{
Quantum machine learning is rapidly emerging as a promising and potentially more sustainable approach that can complement traditional, energy-hungry HPC, AI, and GPU resources—particularly for the demanding global challenges the world urgently needs to address. In recent years, the first prototypes of quantum processors have reached the market, are sometimes available via the cloud, and distributed quantum computing is beginning to move from a largely theoretical concept to a demonstrated reality. However, these devices are still noisy and qubit-limited. Therefore, it is desirable to design quantum machine learning models with a reduced quantum register and a limited number of quantum gates, avoiding overly deep and wide quantum circuits. Here, we explore this idea by proposing a novel divide-et-impera–based training protocol that splits quantum neural networks into smaller components and better governs their optimization. We benchmark our approach on real data and real-world prediction tasks in intertwined energy and environmental domains, such as the status of charging stations for electric vehicles and the global air quality index. Our results suggest a design principle for quantum machine learning algorithms across a wide range of applications, which can be more feasibly tested on state-of-the-art quantum processors or within modular quantum architectures being developed for future hybrid infrastructures interconnected via high-speed communication links.
}


\maketitle

\section{Introduction}

The field of Quantum Computing (QC) has experienced rapid growth in recent years, both in terms of purely theoretical protocols and interest in real-world implementations, driven by the first QC-based approaches outperforming classical solutions. This has led to the development of novel quantum algorithms that, despite being limited by current NISQ hardware, help us understand how QC can improve existing technologies. One of the most promising near-term applications of QC is in Machine Learning (ML), where the QC paradigm is combined with ML training protocols to produce trainable quantum models that can provide greater expressive power while tolerating the noise that affects current Quantum Processing Units (QPUs).
From the union of these disciplines, QML was developed~\citep{tychola_quantum_2023, schuld_introduction_2015, biamonte_quantum_2017}, which has shown promising results in a multitude of fields, e.g., classification~\citep{rebentrost_quantum_2014} and anomaly detection~\citep{bhowmik_quantum_2024}. In these works, it becomes clearer that QML can be used even with our current hardware limitations, especially when used in aid of existing classical models, which are then combined into \emph{hybrid models}. Similar to general QC algorithms, QML solutions have their own set of limitations caused by the noisy implementations.
\begin{enumerate*}[label=(\roman*)]
    \item They are still subject to noise~\citep{resch_benchmarking_2022}, even though the stochastic nature of ML models can contain, or even exploit, the perturbations in order to improve performance~\citep{parigi_quantum-noise-driven_2024}.
    \item The training process of a QML model can be compromised by the presence of \emph{barren plateaus}~\citep{larocca_barren_2025}, which prevent overly complex models from learning from the data.
    \item Training on a large number of parameters can be expensive, due to the training procedure for a quantum model requiring a number of circuit calls that scales linearly with the number of parameters~\citep{wierichs_general_2022}.
\end{enumerate*}
Solving complex tasks requires larger \textit{quantum circuits}, extending the dimension of the state space by using more qubits and performing more complex operations; this results in a build-up of noise that can be avoided only by restricting the number of qubits used, a task that can be accomplished either by reducing the problem dimensionality through feature reduction and feature selection, which may, however, result in excessive information loss, or by limiting the application of quantum algorithms to smaller toy problems. In special cases where the data can be efficiently encoded into the circuit, some interesting results have been achieved in~\citep{kalinin_security_2023, tscharke_semisupervised_2023}; however, these are not general cases and are limited at scale, emphasizing the need for a more general approach.
\cite{pesah_absence_2021} proved that, in Quantum Convolutional Neural Networks, the gradient vanishes at a rate that is at most polynomial, making them more suited for variational training, at the cost of fixing the ansatz structure.

\subsection*{Training with Quantum Neural Networks}\label{limitations}
\begin{figure}[t]
    \centering
    \includegraphics[width=\linewidth]{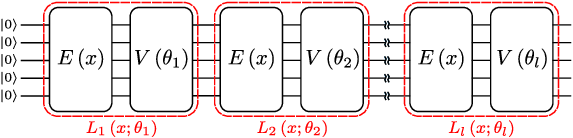}
    \caption{Data re-uploading scheme: similarly to \textit{skip-connections}~\citep{xu_development_2025} in classical NNs, the input is constantly re-uploaded between every variational block.}
    \label{fig:reupload}
\end{figure}

\begin{figure}[t]
    \centering
    \includegraphics[width=\linewidth]{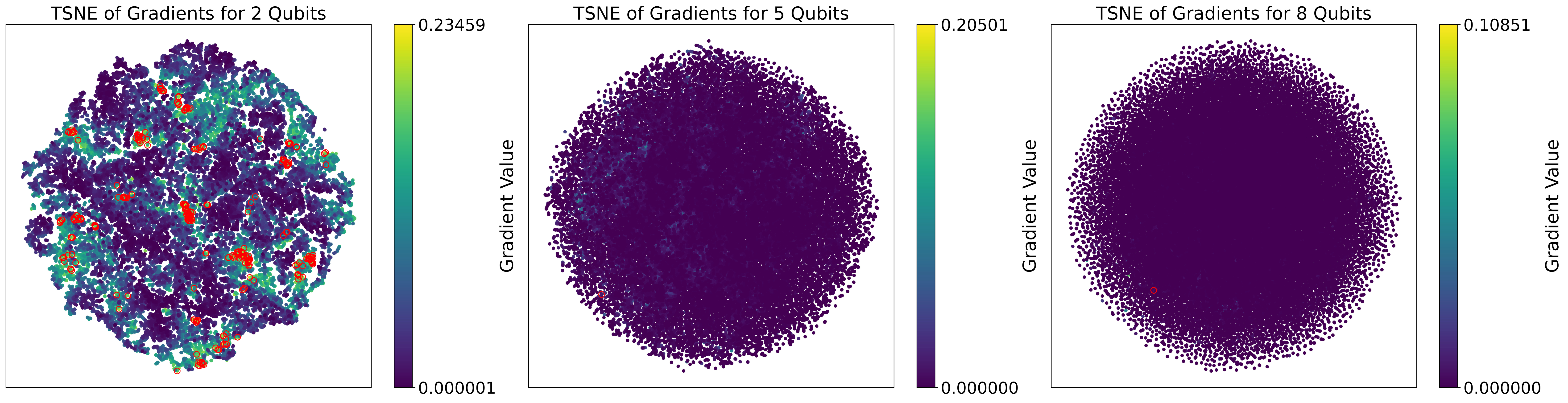}
    \caption{\textit{T-distributed Stochastic Neighbor Embedding} projection of the landscape of the gradients with different circuit sizes, with high values circled in red: with a low qubit count (a), the gradient is well distributed across the landscape, but as the number of qubits increases (b) the gradient starts to concentrate in  smaller region, and at high qubits count (c), the gradient can be assumed null almost everywhere.}
    \label{fig:bp}
\end{figure}

Hybrid ML models combine the flexibility of classical ML with the expressive power of variational quantum circuits (VQCs), enabling the representation of more complex data distributions or embeddings. A VQC is a quantum circuit that presents a set of learnable parametric gates that are associated with a family of quantum functions, and using a VQC it is possible to define the simplest form of Quantum Neural Network (QNN), which consists of three operations: \begin{enumerate*}[label=(\roman*)]
    \item encode the data, defining a linear map $E(x)$;
    \item apply the parametrized operation $V(\theta)$;
    \item measure the expectation value of some observable $\mathcal{O}$.
\end{enumerate*}
The first step allows mapping our data into the quantum state space using different strategies, each of them requiring different circuit size and circuit depth to implement, while the second step makes the quantum layer capable of learning, through the additional degrees of freedom provided by the tunable parameters, and, lastly, the measurement provides the output of the layer, which can be directly used to optimize a loss function by means of the gradient descent algorithm.
At this level of abstraction, as long as the gradient of the circuit’s parametrized elements can be computed, there is no essential distinction between a classical and a quantum layer, as both can be optimized using analogous methods.
State-of-the-art QML techniques define the VQC as a linear function
\[\mathcal{U}\left(x; \theta\right)=\prod_{i=1}^{R}V_i\left(\theta_i\right)E\left(x\right)\] where ${E}\left(x\right)$ is the encoding function that maps the input data into the quantum state $\ket{x}$ and $\left\{V_i\left(\theta_i\right)\right\}_i$ is the set of parametrized unitary matrices, each with its own parameters. This formulation, visualized in Figure~\ref{fig:reupload}, is known as the data re-uploading scheme~\citep{reupload_P_rez_Salinas_2020}, which improves circuit expressivity, at the cost of increasing the amount of interaction present inside the system, which leads to a flatter loss landscape, that makes the training difficult for the derivative-based methods, as the number of circuit repetitions $R$ becomes large~\citep{mcclean_barren_2018}. 
Thus, training a quantum layer involves learning a parametrized quantum function \[\psi(x;\theta) = \bra{0}\mathcal{U}(x;\theta)^\dagger\mathcal{O}\mathcal{U}(x;\theta)\ket{0}\] typically optimized via gradient descent. However, despite the efficiency of gradient descent for classical neural networks—where gradients are cheap to compute—it poses challenges in the quantum setting: \begin{enumerate*}[label=(\roman*)]
    \item Using the parameter-shift rule, each partial derivative requires multiple circuit evaluations, resulting in a training cost proportional to the number of trainable parameters, which limits its scaling capabilities.
    \item As shown in Figure~\ref{fig:bp}, increasing the number of qubits causes the gradients to exponentially concentrate in narrow regions and vanish elsewhere, making gradient-based updates ineffective.
\end{enumerate*}
This flattening out of the function $\psi(x;\theta)$ into so-called barren plateaus is a well-known problem across the field of variational quantum algorithms, for which no straightforward solutions are currently available.

Several solutions have been proposed to address these challenges, though they often rely on assumptions about the circuit’s ansatz~\citep{coyle_training-efficient_2025}, limiting generality, or they reduce training costs through approximate gradient computation~\citep{gidi_stochastic_2023}, which may not always suit the problem at hand. Rather than limiting the range of applications, we build up from standard state-of-the-art models, which mitigate the barren-plateau effect by using \textit{local observables}~\citep{cerezo_cost_2021} and thus measuring each qubit individually. This sacrifices some generality but causes the gradient to vanish polynomially rather than exponentially for shallow circuits.
We combine these solutions with the design of a Hybrid Neural Network that employs both classical and variational layers, using the flexibility of classical networks to overcome limitations in both the mathematical construction and the implementation of quantum layers, while still improving prediction performance in the right circumstances. A prominent example is the work of Huang et al. (2024), \textit{Post-variational quantum neural networks}, where the authors provide a hybrid architecture that addresses the trainability issues of the quantum layer by shifting all trainable parameters to the classical layers; this learns from quantum-processed data a better decision boundary than classical layers alone. Hybrid QNNs have found successful applications in different settings, including image classification~\citep{senokosov_quantum_2024, hafeez_h-qnn_2024} and drug development~\citep{sagingalieva_hybrid_2023, anoshin_hybrid_2024}. 
Despite promising results, the trainability issue still hinders the potential of these models to exploit the possibility of learning a quantum function alongside a classical one, reducing the adaptability of the model. 
Motivated by the lack of a method capable of dealing with variational training in a scalable way, we propose a divide-et-impera approach that allows the reformulation of the training into smaller subroutines, allowing us to keep under control the scale of each quantum process, reducing the amount of resources required for carrying out the computation without restricting the class of executable quantum circuits. 
The context of quantum divide-et-impera has already been explored in some relevant work. For example, \cite{childs_quantum_2025} presented a general divide-and-conquer framework alongside an analysis of a general cost reduction of a set of decision problems. The \textit{Quantum Divide and Compute} method~\citep{ayral_quantum_2020, ayral_quantum_2021} is based on dividing the circuit execution into independent chunks, allowing a quantum algorithm to be executed using fewer qubits, at the cost of a number of circuit calls exponential in the number of circuit cuts. A relevant application of the divide-et-impera concept in variational training is the \textit{Quantum Divide and Conquer Algorithm} proposed by~\cite{tomesh_divide_2023}: in that paper, the authors successfully map a single instance of the Maximum Independent Set problem into multiple smaller subproblems, individually tackled by different processors. Another promising application has been proposed by~\cite{fang_divide-and-conquer_2024}, who used a problem-informed ansatz and a decomposition of the Hamiltonian to solve a Variational Quantum Eigensolver based genome assembly problem up to 140 qubits using a simulation with only 10 qubits.

Our work contributes to the literature, which, to the best of our knowledge, lacks such methods in the form of a divide-et-impera training scheme for Hybrid QNNs that can be generally applied to any scenario, providing a reduction in training costs and hardware requirements without incurring degradation of the model performance by acting on a data level. 

In this work, we contribute to the literature which, to the best of our knowledge, lacks such methods by proposing a divide-et-impera strategy to train a hybrid neural network model on larger problems by decomposing the task into smaller components. This allows the quantum computation to be distributed across multiple QPU calls, achieving appealing performance while reducing the resources needed for training. The generality of the proposed approach allows its application to a wide range of settings, providing a reduction in training costs and hardware requirements without incurring degradation of model performance by acting at the data level. 
The article is organized as follows: in Section~\ref{sec:results}, we showcase the results of our approach on two different problems and discuss the method’s effectiveness when features are not directly decomposable. Finally, in Section~\ref{sec:methods}, we provide a comprehensive description of our method, along with the data and experimental setup.

\section{Methods}\label{sec:methods}

\subsection*{Prediction Using Quantum Embeddings}\label{method}
\begin{figure}[b]
    \centering
    \includegraphics[width=\linewidth]{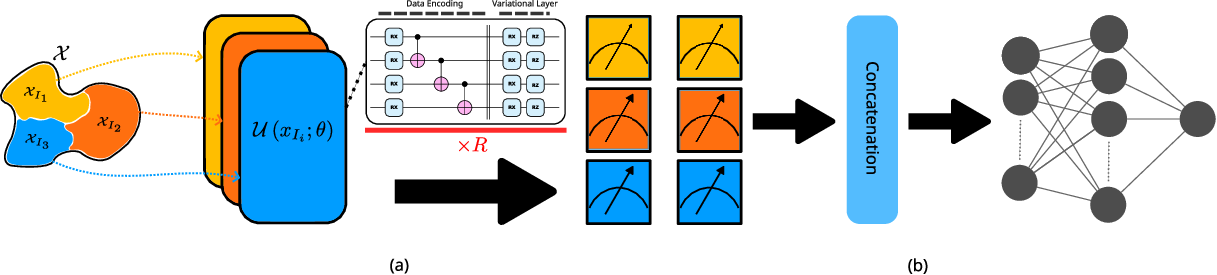}
    \caption{Decomposition scheme of the proposed model. For each set of features a dedicated VQC creates an output representation (a). The embeddings computed by each circuit are concatenated and used by a classical network in order to obtain the final output of the layer (b).}\label{fig:scheme}
\end{figure}

Inspired by common patterns in algorithm design, and by the lack of similar techniques in the context of hybrid QNNs, we propose a divide-et-impera--based solution for the variational training of a QNN. By decomposing an initial $Q$-sized circuit into $k$ sub-circuits of size $q_i < Q$, $i=1,\dots,k$, it is possible to significantly reduce the resources required for the computation, at the cost of a partial loss of information due to the reduced circuit size. This loss is mitigated by using a reconstructing network, which combines the outputs of all VQCs to produce the final predictions. 
We propose a model construction scheme that takes into account these limitations, by processing only a subset of the input at a time, using different parametrized circuits for each component, resulting in a set of outputs that will be concatenated and then used as the input of a successive layer, summarized in Figure~\ref{fig:scheme}. This follows the intuition about classical Deep NNs learning internal representations of the feature space $\mathcal{X}$ inside the hidden layers of the model: by providing a separation at the feature level, it is possible to learn a meaningful quantum representation of the subsets provided, as long as the features are related to each other. Formally, given an input vector $x\in\mathbb{R}^N$ and a set of indices $\left\{I_j\right\}_j$ such that $\bigcup_j I_j = \left\{1,\ldots,N\right\}$ we define the output of the quantum model as \[z = \odot_j VQC_j(x_{I_j};\theta_j),\] where $x_{I_j}$ is the subset of the input specified by the indices in $I_j$, and $\odot$ is the concatenation operator. The resulting output $z \in \mathbb{R}^M$, with dimension $M=\sum_j \left|I_j\right|$, represents the concatenation of the \textit{quantum embeddings} learned using the VQCs, and a final stitching shallow neural network learns to use the learned embeddings to perform a prediction \[\hat{y}=MLP(z;\theta_{pred}),\] with $MLP$ denoting a classical neural network defined using the perceptron model, as in~\cite{mlp_def}.
Each circuit can have an independent design; however, in order to simplify the analysis of the results all circuits, unless otherwise stated, share the same ansatz (reported in Figure~\ref{fig:ansatz}), with \textit{angle encoding} \[
\ket{x} = \otimes_{k=1}^d  \left(\cos\left(\frac{x_k}{2}\right) \ket{0} - i \sin\left(\frac{x_k}{2}\right) \ket{1}\right),
\] and 2-angle rotations as a variational layer, in addition to a set of CNOT \textit{nearest-neighbor} connections to introduce entanglement.

This scheme, summarized in Figure~\ref{fig:scheme}, allows the quantum computation to be distributed into smaller circuits that are less influenced by noise and cheaper to run, since they require fewer gates to operate, with only a small overhead in the number of total parameters required. In particular, when all sub-circuits use the same regular ansatz (i.e., identical operations are applied to each qubit, as in Figure~\ref{fig:ansatz}) and the index sets $\left\{I_j\right\}_j$ form a partition of the feature space (no overlaps), the total number of trainable parameters is unchanged: the parameters are simply redistributed across circuits. The embeddings produced by the VQCs are then combined into a single vector, which is then processed by a final shallow classical layer, providing the adequate output format. 

\begin{figure}[t]
    \centering
    \includegraphics[width=0.5\linewidth]{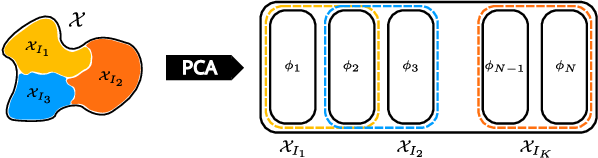}
    \caption{After processing the data, the subset can be retrieved using a sliding window with an arbitrary window size and step.}\label{fig:pca-preprocess}
\end{figure}

In cases where it is not possible to manually provide a separation of the feature space $\mathcal{X}$ (e.g. when the feature space is too large or the individual features are not meaningful without the others for context), we can enforce the separation by further preprocessing the dataset. This can be achieved using \textit{Principal Component Analysis}~\citep{ringner_what_2008} to find a linear map $F$ that can be used to express our data into a new space, where every feature is independent of the others. In this way, every subset of $\Phi=F\mathcal{X}$ can be chosen as a candidate. Additionally, the Principal Component Analysis (PCA) provides a natural ordering of this new feature space given by the explained variance of the new features, allowing selection of suitable subsets using a sliding-window approach, as depicted in Figure~\ref{fig:pca-preprocess}.

We tested our method on two different real-world problems: one provides a more accessible setting, with a straightforward criterion for separating the data, while the other represents a more general case in which the problem cannot be decomposed directly.
We used a hybrid architecture, comprising a quantum layer that aims at learning a representation of the input data, followed by a classical feedforward layer, which creates the actual prediction of the station state. The computational cost of the quantum layer is reduced by distributing the data across multiple VQCs, each processing a subset of the input $x_{I_i} \in x$, and then aggregating the output into a single vector, as shown in Figure~\ref{fig:scheme}. Each VQC can have a different ansatz, however if not stated otherwise, they employ a simple structure reported in Figure~\ref{fig:ansatz}: the data is encoded using the angle encoding, with entanglement introduced with a chain of $CX$ gates, with two variational rotations per qubit. Following the results in \cite{reupload_P_rez_Salinas_2020}, the circuits follow the \emph{data re-uploading} scheme, depicted in Figure~\ref{fig:reupload}, and at the end of the computation, the expectation value $\left<Z\right>$ of each qubit is measured. The outcomes of all the circuits are then concatenated and fed into a classical layer to produce the final prediction.
\subsection{Dataset and Experiment description}
The proposed architecture is applied to the following problems, which can be accessed through the Data Availability section.
\subsubsection*{Charging Stations}
The dataset was originally provided by Amara-Ouali et al (2023) - Forecasting Electric Vehicle Charging Station Occupancy: Smarter Mobility
Data Challenge - where the authors organized a hackathon to predict the state of 91 electric vehicles (EV) charging stations distributed in the north, south, west and east areas of Paris, denoted by $\mathcal{A} = \left\{N, S, W, E\right\}$. Each charging station has three plugs, which can be in any of the states \emph{available}, \emph{charging}, \emph{passive}, and \emph{other}. Therefore, the state of each station can be described by a vector of integers $s = \left(a, c, p, o\right)$, in which each component represents the number of plugs in the corresponding state, with the constraint $a+c+p+o=3$.
The features are a sequence of samples spanning from 2020--03--07 00:00 to 2021--02--18 23:45 CET obtained every 15 minutes, containing the information about the station states and locations (coordinates, ZIP code and area designation), with additional time information. The last two samples are used to predict the next state of the station.

\begin{figure}[b]
    \centering
    \includegraphics[width=\linewidth]{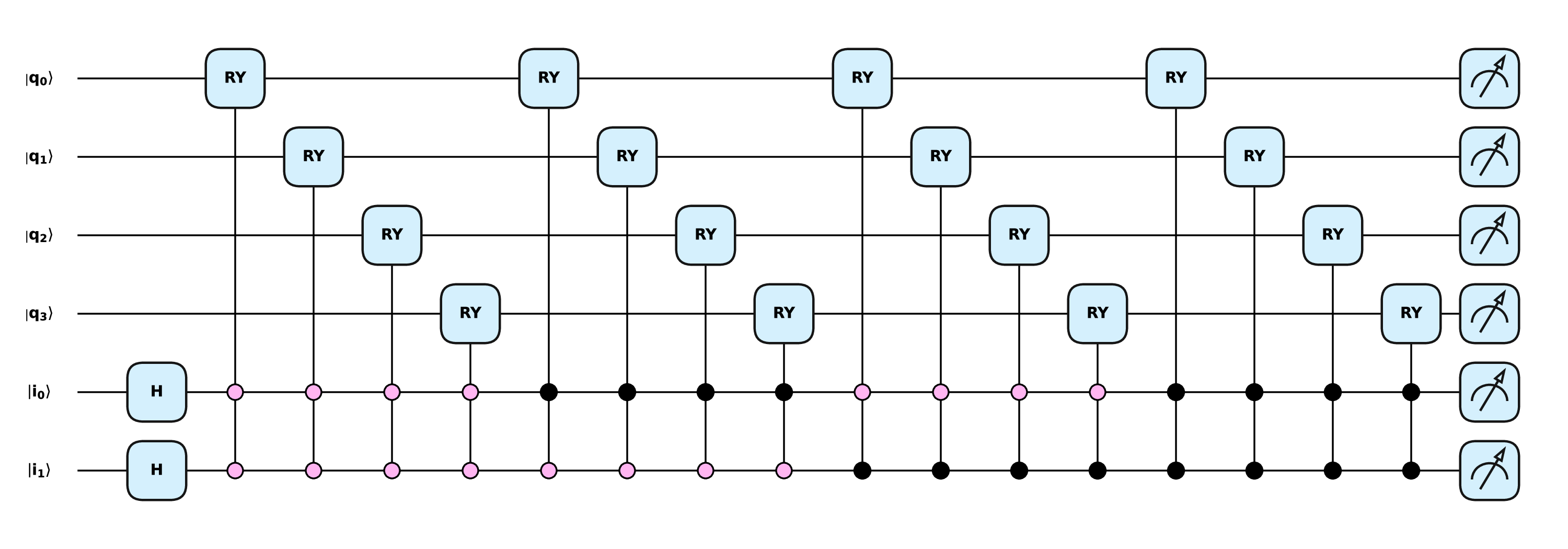}
    \caption{Ansatz used for processing the sequence of states, which uses an additional register to encode the sequence index.}
    \label{fig:frqs-ansatz}
\end{figure}

The main proposed split of the feature set is the partition given by $I_1 = \left\{state(t)\right\}$, $I_2 = \left\{state(t-1)\right\}$, $I_3 = \left\{ToD, DoW, trend\right\}$ and $I_4 = \left\{Lat, Lon, Area, ZIP\;Code\right\}$: since all of these sets are meaningful on their own, we can avoid overlapping sets, without overhead in the parameters needed for the VQCs. This solution is tested against two classical models: a sequential model, which processes the time information using a transformer-based encoder~\citep{vaswani_attention_2017}, and a classical model equivalent to the hybrid model described above, in terms of latent space dimensions, as indicated in~\citep{inflate-Sakhnenko_2022}, where the same divide-et-impera data decomposition approach is applied. 
Lastly, we compare the results of the quantum model at inference time in the presence of noise, validating its performance in a more realistic setting. Noise is simulated using a \emph{depolarizing channel}, with one-qubit and two-qubit gate error probabilities aligned with the value provided by IBM QPUs.
Furthermore, noiseless results are provided for the decomposition strategy $I_1 = \left\{state(t-1), state(t)\right\}$, $I_2 = \left\{ToD, DoW, trend\right\}$ and $I_3 = \left\{Lat, Lon, Area, ZIP\;Code\right\}$, with $VQC_1$ using the same ansatz structure but a wider register size, referred to as Hybrid Model \emph{Unified State} (US), and with a sequence-sensitive ansatz named Hybrid Model \emph{Sequential} (Seq), reported in Figure~\ref{fig:frqs-ansatz}. This ansatz can handle sequences of size $L$ using only $4 + \lceil \log_2 L\rceil$ elements instead of the $4 \times L$ scaling of the initial solution. 

The goal is to perform the prediction of the states at 3 different levels of granularity: the station level, the area level and the city level. The result is expressed using the $L_1$ norm as a scoring function \[
    \mathcal{S}\left(y, \hat{y}\right) =  \sum_{i=1}^{91} \|y_i - \hat{y}_i\|_1 + \sum_{a \in \mathcal{A}} \|y_a - \hat{y}_a\|_1 + \|y_g - \hat{y}_g\|_1,
\] which can be interpreted as the total error at all levels of precision in all stations at a given time, from which it is possible to retrieve single station and area information.

\begin{figure}[t]
    \centering
    \includegraphics[width=0.7\linewidth]{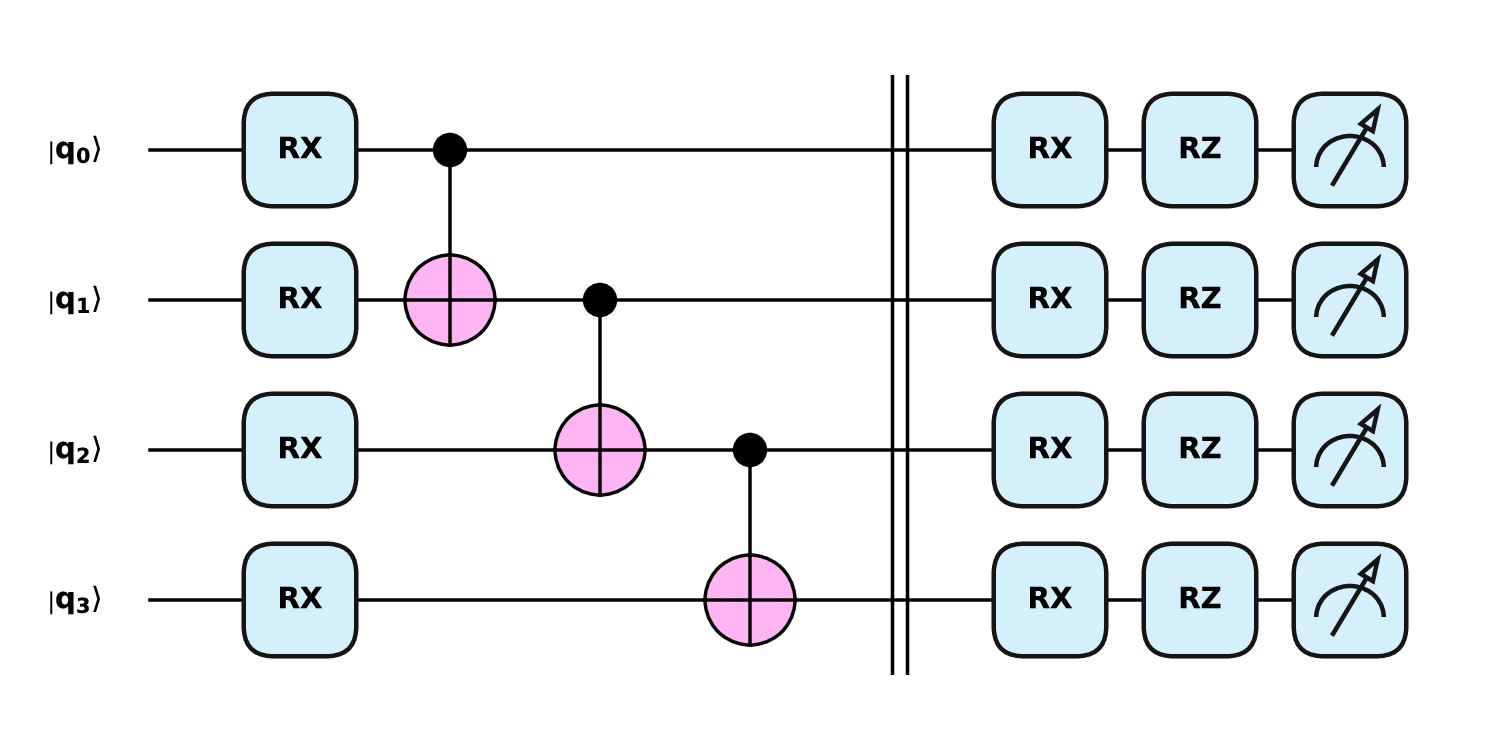}
    \caption{Circuit that computes the map $\ket{\phi(x; \theta)}= U(x;\theta)\ket{0}$}\label{fig:ansatz}
\end{figure}

The data is divided into two non-overlapping training and test sets, which span different timeframes. The training set is further split into a training set and a validation set, comprising different stations.   
Since the scoring function $S$ is not differentiable, it cannot be optimized directly. Therefore, we define a surrogate loss function \begin{gather*} 
    \mathcal{L}\left(y, \hat{y}\right) = \mathcal{L}_{S}\left(y, \hat{y}\right) + \mathcal{L}_{A}\left(y, \hat{y}\right) + \mathcal{L}_{G}\left(y, \hat{y}\right) + \alpha \mathcal{L}_{P}\left(\hat{y}\right), \\
    \mathcal{L}_{S}\left(y, \hat{y}\right) = \sum_{i=1}^{91} \|y_i - \hat{y}_i\|_2, \\
    \mathcal{L}_{A}\left(y, \hat{y}\right) = \sum_{a \in \mathcal{A}} \|y_a - \hat{y}_a\|_2, \\ 
    \mathcal{L}_{G}\left(y, \hat{y}\right) = \|y_g - \hat{y}_g\|_2, \\ 
    \mathcal{L}_{P}\left(\hat{y}\right) = \sum_{i=1}^{91}\left[\left(\sum_{j} y_{i,j}\right)-3\right]^2
    \end{gather*}
where the new term $\mathcal{L}_P$ is added in order to enforce the constraint $a+c+p+o=3$, preventing the model from generating invalid states. The relevance of this penalty term is determined by the hyperparameter $\alpha$, which will be set to $1$ and omitted henceforth, since this value was found to be adequate.

The experiments in this task are performed in both noiseless and noisy regimes. After training, we introduce quantum noise into the model with the goal of evaluating its resilience to perturbations that are inevitably introduced when applying this method to a real QPU. The noisy experiments are conducted with two quantum channels
\begin{gather*}
    \mathcal{E}_\text{bit-flip}(\rho) = p X\rho X^\dagger + \left(1-p\right)\rho \quad X = \begin{bmatrix} 0 & 1 \\ 1 & 0 \end{bmatrix} \\
    \mathcal{E}_\text{depol}(\rho) = p \frac{I}{2} + \left(1-p\right) \rho \quad I = \begin{bmatrix} 1 & 0 \\ 0 & 1 \end{bmatrix}
\end{gather*} that apply a perturbation to the quantum state after each operation with probability $p$, tuned using values related to real QPUs.
\subsubsection*{Global Air Quality}

The Air Quality dataset consists of environmental measurements of air quality indicators across six cities in the world, sampled during the year 2023, with one sample per hour. The sample contains the sampling time, the concentration of Carbon Monoxide (CO), Carbon Dioxide (CO$_2$), Nitrogen Dioxide (NO$_2$), Sulfur Dioxide (SO$_2$), Ozone Concentration (O$_3$), small and large Particulate Matter (PM2.5) and (PM10) elements in the air, with the goal of predicting the European Air Quality Index (AQI). A splitting of the dataset into training, validation, and test sets is obtained by randomly selecting different samples for each individual city.
Given the high complexity of the data, finding a clear division of the dataset is not possible: therefore we apply the PCA preprocessing described in Section~\ref{method}, testing circuit sizes from 3 to 6 qubits at different step sizes. The results are compared with a standard MLP model and a standard QNN with a shallow classic predictor. The ansatz used in both our solution and the quantum baseline share the same design with the previous experiment (Figure~\ref{fig:ansatz}). The models are trained to minimize the AQI mean square error, and are compared using mean absolute error. Therefore, the scoring function and the loss in this case are given by
\begin{gather*}
    \mathcal{S}(y, \hat{y}) = \frac{1}{N}\sum_{i=1}^{N}\left|y_i - \hat{y}_i\right|, \quad
    \mathcal{L}(y, \hat{y}) = \frac{1}{N}\sum_{i=1}^{N}\left|y_i - \hat{y}_i\right|^2,
\end{gather*}
respectively.

\subsubsection*{Experiment Description}
Both experiments share a similar setup: both implementations were performed using PennyLane (Bergholm et al., 2022, \textit{PennyLane: Automatic differentiation of hybrid
quantum-classical computations}) and PyTorch~\citep{paszke_pytorch_2019}. The models are trained using the Adam optimizer with an initial learning rate $\gamma_0=10^{-3}$, with a linear scheduler with initial and final factors of $(1, 10^{-4})$. The \textit{charging station} and \textit{air quality} problems are trained for $30$ and $50$ epochs, respectively.

For the \textit{charging stations} dataset, we assign $\frac{2}{3}$ of the provided training dataset to the training split, and the rest to the validation split: the test split is provided in a dedicated file. For the \textit{air quality} dataset, for every city, the training split makes up to $60\%$ of the provided data, with the remaining samples evenly distributed between validation and test split.

\section{Results}\label{sec:results}
To validate the efficacy of the proposed divide-et-impera architecture, we evaluate our model across the two aforementioned scenarios: the electric vehicle charging station state prediction and the global air quality index forecasting, and compare their results with benchmark models. 


\subsection*{Charging Stations}
As we can see in Table~\ref{table:cs-result-parameters}, the base hybrid model outperforms both classical implementations while using significantly fewer parameters, reducing the possibility of overfitting the dataset and improving the overall generalization capabilities of the model in a case where simulated training using a single circuit to handle all the data would not be possible. 
Notably, the base hybrid model demonstrates superior performance compared to both the sequence-sensitive and unified state variants, suggesting that strictly isolating the temporal information across distinct circuits and delegating its interpretation to the classical part of the model yields a more resilient representation than processing it within a single quantum circuit.
Moreover, as reported in Table~\ref{table:cs-result-parameters}, decomposing the data yields results comparable with the classical models, and the quantum based solution generally exhibits a lower variance in predictions, leading to more stable and reliable behavior, even in the presence of noise, although a systematic, noise-dependent reduction in performance was expected and observed. 
Though not as effective in training the model in the presence of noise, this strategy still provides some degree of error mitigation, putting a lower bound on overall performance. In noisy regime, as expected, we observe a degradation in performance, that is however contained: given a starting score of $53.86$ for the noiseless model, employing a depolarizing noise channel we still get a score of $60.8$, while when using a bitflip channel we get a worse score of $68$, both of them still comparable to noiseless and classic solutions.

\begin{figure}[t]
    \centering
    \includegraphics[width=\linewidth]{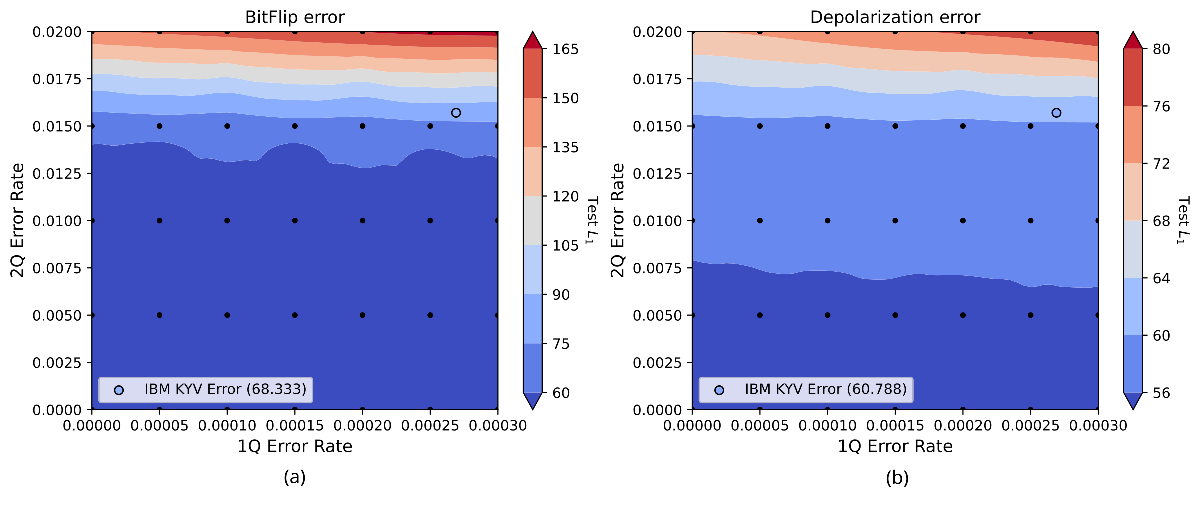}
    \caption{Test score response with the Bit-Flip (a) and Depolarization (b) error models, as a function of the incidence rate of single- and two-qubit operations. The heatmap is obtained through cubic interpolation of the values computed at the marked points. The empty circles denote the error rate of the IBM\_KYIV QPU.}\label{fig:noise-cs}
    
\end{figure}

A more detailed analysis of the noise is provided in Figure~\ref{fig:noise-cs}, which reports the score of the model at different error rates for one-qubit and two-qubit gates, providing an insight into the major incidence of the error of the two qubit gates, that defines a general error cut-off region.
In Figure~\ref{fig:noise-cs}, a more comprehensive noise response of the score is provided, showing how destructive the build-up of noise can be at different intensities, including error rates from actual QPUs. In some simpler noise models with a slightly lower error rate, we would achieve performances comparable to an ideal quantum training.

\subsection*{Air Quality}

From the results we can observe how both quantum-based solutions fall behind classic solutions; however, as can be noted in Table~\ref{tab:airq-results}, preprocessing the data using PCA seems to be an effective way to enable the data decomposition approach to be applied in more general settings, granting this method a broader spectrum of possible applications, at the expense of a small preprocessing overhead.
\subsection*{Discussion}
We showed that QML can be a more than viable solution to some categories of real-world problems, providing desirable properties and competitive performance. We proposed a data-driven divide-et-impera method to decompose a training problem into smaller components, allowing it to be processed using smaller modules. 
Building up from the flexibility of Hybrid QNNs, this approach allows us to tackle tasks of greater size with the same hardware availability, retaining or improving performance concerning standard hybrid solutions with almost no overhead in terms of the number of parameters employed. The solution remains viable even in settings where the data decomposition cannot be directly applied, using PCA as an additional preprocessing step, making this method generally applicable to a multitude of scenarios. Moreover, by reducing the size of the qubit register, we implicitly reduce the circuit depth required by standard ansatz constructions, with a potential impact on noise buildup. 

Delegating the computation to different sub-circuits not only reduces the required computational power but also allows the computation to be performed in parallel while retaining resource efficiency. This is particularly beneficial given the generally super-linear scaling with problem size, further testifying to the advantage of performing distributed inference using a small fraction of parameters compared to classical infrastructure. 
A possible development of this work could be directed towards dealing with independent noise models for each circuit, working in a regime of Distributed Quantum Machine Learning, assessing the capacity of different existing ML solutions to learn to unify mismatching noise patterns, and enabling different QPUs to cooperate.

\backmatter

\bmhead{Competing Interests}
The authors declare no competing interests.

\bmhead{Acknowledgments}
This study received no funding.

\bmhead{Data Availability}
The data used in this article can be retrieved from the following websites: \begin{itemize}
    \item EV Stations dataset: \url{https://gitlab.com/smarter-mobility-data-challenge/tutorials}
    \item Air Quality dataset: \url{https://www.kaggle.com/datasets/youssefelebiary/global-air-quality-2023-6-cities?resource=download}
    \item IBM QPU data: \url{https://quantum.cloud.ibm.com/computers}
\end{itemize}
\bmhead{Code Availability}
The underlying code for this study is not publicly available but may be made available to qualified researchers upon reasonable request from the corresponding author. 
\bmhead{Author Contributions}
E.C. carried out the algorithm implementation and experiments. E.C. and F.C. conceived the methodology, while A.A. and F.C. proposed and supervised the project. All authors contributed to the discussion, analysis of the results, and the writing of the manuscript.

\begin{table}[h]
    \centering
    \begin{tabular}{ l c c r  }
        \toprule
        Model & Score (\textbf{$\downarrow$}) & std. dev. & Parameters \\
        \midrule
        Sequential &  57.58 & 4.6 & $\approx 3\cdot10^6$ \\ 
        Classical & 65.03 & 4.5 & $\approx 3\cdot10^5$\\
        Hybrid Model & \textbf{53.86} & 0.76 & $\mathbf{\approx 4\cdot10^4}$ \\
        Hybrid Model (US) & 60.13 & 1.9 & $\mathbf{\approx 4\cdot10^4}$ \\
        Hybrid Model (SEQ) & 55.94 & 0.8 &  $\mathbf{\approx 4\cdot10^4}$ \\
        \bottomrule
    \end{tabular}
    \caption{These are the results in noiseless settings on the Charging Stations dataset. The statistics are computed using the results of 20 runs. }
    \label{table:cs-result-parameters}
\end{table}

\begin{table}[h]
\centering

    \begin{tabular}{ll cccccc} 
\toprule
 & & \multicolumn{6}{c}{Score (\textbf{$\downarrow$})}  \\
Model & Parameters & Dubai & Sydney & New York & London & Brasilia & Cairo \\
\midrule
Classic & $\approx 1\cdot10^6$ & \textbf{9.73175} &\textbf{ 4.87031 }&\textbf{ 8.34855} & \textbf{3.27224} & \textbf{2.09935} & \textbf{9.06304} \\
            Ours & $\approx 1.8\cdot10^3$ & 12.9519 & 5.95415 & 10.46712 & 5.0847 & 5.24563 & 9.74079 \\
            Hybrid & $\mathbf{\approx 1.4\cdot10^3}$ & 15.02054 & 7.19897 & 11.95646 & 6.66119 & 6.36074 & 10.66098 \\
\bottomrule
\end{tabular}

    \caption{Results on the Air Quality, divided by the target city: even though the classic model performs better than both hybrid models, the results show how separating the data brings consistent improvements even when the initial assumptions are not met.}
\label{tab:airq-results}
\end{table}
\bibliography{sn-bibliography}

\end{document}